\documentclass[conference]{IEEEtran}
\IEEEoverridecommandlockouts
\usepackage{cite}
\usepackage{amsmath,amssymb,amsfonts}
\usepackage{algorithmic}
\usepackage{graphicx}
\usepackage{textcomp}
\usepackage{xcolor}
\usepackage{float}
\usepackage[hidelinks]{hyperref}

\def\BibTeX{{\rm B\kern-.05em{\sc i\ke rn-.025em b}\kern-.08em
    T\kern-.1667em\lower.7ex\hbox{E}\kern-.125emX}}
\begin{document}

\title{Electrical Power Grid Reliability Assessment Exposed to Straight Line Winds\\ }

\author{\IEEEauthorblockN{Satish Adhikari}
\IEEEauthorblockA{\textit{Electrical Engineering \& Computer Science} \\
\textit{University of Wyoming}\\
Laramie, United States \\
sadhika3@uwyo.edu}
\and
\IEEEauthorblockN{Md Abul Hasnat}
\IEEEauthorblockA{\textit{Electrical Engineering \& Computer Science} \\
\textit{University of Wyoming}\\
Laramie, United States \\
mhasnat@uwyo.edu}
}

\maketitle

\begin{abstract}
Maintaining the resilience and reliability of electrical power grids under weather driven disturbances is a critical challenge for modern power system planning and operation. Among weather driven hazards, tornadoes and hurricanes are mostly considered  for research purpose overlooking straight line winds despite its significant contribution to transmission line outages and cascading effects. Accurate modeling of these events is essential for operational decision making, and long term infrastructure planning and investment. This research explores the probabilistic risk assessment of transmission lines under straight line winds and their impacts on system level performance. The approach examines the probability of transmission towers failure, transmission lines outages, cascading effects, and identify critical transmission lines. The IEEE 118 bus system was utilized to validate the proposed approach and to demonstrate how wind induced component failures propagate throughout the network, affecting power grid reliability. 

\end{abstract}


\section{Introduction}

\subsection{Motivation}

Power system studies under extreme weather events have gained increasing attention in recent research. The increasing frequency and severity of weather driven disturbances have resulted in widespread outages and significant economic and social impacts \cite{Bhusal2020_Resilience}. In recent years,  such disturbances have caused over 80\% of major power outages threatening reliable grid operation \cite{NWS_HighWind,Chaudhary2025_Storage}. Since electrical power system is highly interconnected and spatially distributed, failures triggered by extreme weather can propagate rapidly, leading to cascading outages and large scale service disruptions. Consequently, understanding and mitigating weather induced risks has become a critical priority for modern grid planning and operation.

Among various weather driven hazards, extreme straight line winds pose a major threat to transmission systems \cite{Greco2025_WindRisk}. These winds are strong, non-rotating winds that move horizontally in a relatively uniform direction. Unlike hurricanes, which are large rotating tropical systems, or tornadoes, which involve intense localized rotation, straight line winds often affect large geographic areas with sustained high wind speeds, exerting substantial mechanical stress on transmission towers and conductors. For example, on December 22, 1982, a severe straight line wind event on the U.S. West Coast caused the collapse of a 500 kV transmission tower, which subsequently impacted an adjacent parallel 500 kV line leading to a large scale outage affecting more than five million customers \cite{McCarthy1996_TowerFailure}. Similarly, in August 2020, a large scale derecho (i.e., a long lived and fast moving band of severe thunderstorms producing widespread straight line winds) impacted the U.S. Midwest, causing simultaneous power outages affecting approximately 1.4 million customers \cite{Dikshit2023_Fragility}. These events demonstrate that straight line winds can trigger transmission line outages, tower failures, and the loss of critical transmission corridors, which may initiate cascading failures across the network.

Straight line wind phenomena such as microbursts, macrobursts, and derechos are relevant hazards due to the region's frequent high wind conditions, particularly, in the context of Wyoming \cite{NWS_HighWind}. Despite their occurrence and potential impact, straight line wind events have received comparatively less attention in power system studies than hurricanes, floods, or earthquakes.  These reasons motivates us to work on a probabilistic assessment framework that links spatial wind exposure, fragility informed component failures, cascading dynamics and reliability analysis to analyze the grid level impacts and identify critical transmission lines.

\subsection{Related Works}
Existing studies on weather driven transmission risk can be classified into component level wind vulnerability modeling and system level assessment. At the component level, prior work has investigated the structural response of transmission towers and conductors under wind loading, including buffeting, galloping, and dynamic response under synoptic and non-synoptic winds \cite{Aboshosha2016_Buffeting}. High fidelity finite element model (FEM) and moment matching methods have also been used to develop fragility curves for transmission towers subjected to straight line winds \cite{Dikshit2023_Fragility}. In addition, probabilistic formulations have been proposed to quantify wind and ice induced failure probabilities of transmission line components \cite{Kudzys2006_WindIce,Panteli2016_Fragility}. These studies provide important insight into component vulnerability; however, they are highly dependent on detailed structural dimensions, material properties also computationally expensive and may not be practical for bulk transmission grid.

At the system level, in recent year, transmission system operators (TSOs) worldwide are increasingly adopting probabilistic risk assessment tools to evaluate the impacts of natural hazards on grid operations, supplementing traditional deterministic approaches based on N-k security criteria \cite{Bompard2013_Threats}.  Recent studies have also examined wind induced line outage probabilities and cascading failure mechanisms in transmission networks \cite{chaudhary2025cascading}. However, this research does not extend their work from wind induced outage analysis to post cascade reliability analysis for identifying wind vulnerable and critical transmission lines.

\subsection{Research Gaps}

Despite ongoing advances in this field of research, several gaps have been identified in the literature. First, the  existing works mostly remain confined to component specific fragility modeling (mostly structural modeling of transmission towers and its fragility analysis) and does not readily scale to grid level assessment.
Moreover, most works on the reliability against extreme wind events primarily emphasis on hurricanes and tornadoes overlooking straight lines winds. Although such studies are valuable for understanding worst case impacts, they may not fully capture the cumulative exposure of transmission networks to spatially distributed straight line wind conditions and its impacts on wind induced outage scenarios.

\subsection{Contributions}
Motivated by the above identified research gaps, this study proposes a probabilistic wind risk assessment framework that evaluates wind induced transmission component vulnerability and its resulting grid level impacts. The main contributions of this articles are as follows:

\begin{itemize}
    \item This paper presents a framework for modeling the uncertainty of spatially distributed straight line winds profile over Wyoming and its risk assessment to transmission line infrastructure.
    \item The research further extends component level fragility evaluation to grid level assessment with cascading failure and reliability analysis to identify critical transmission lines and support grid planning, operation, and reinforcement decisions. 
\end{itemize}

The remainder of this paper is organized as follows: Section II details the proposed risk assessment methodology. Section III presents the simulation setup and analyzes the corresponding results. Section IV summarizes the key findings and discusses potential directions for future work.

\section{Methodology}

This section provides the details of proposed methodology to assess the risk of transmission systems exposed to extreme straight line wind events. The framework integrates wind hazard modeling, component level fragility assessment, and transmission line failure probability analysis to quantify system level impacts.

\subsection{Wind Exposure Mapping}
The wind exposure field can be developed from geographically distributed wind observation stations to represent the spatial variability of straight line wind exposure. Let the set of wind observation stations be denoted by $\mathcal{S}=\{(\lambda_i,\phi_i,v_i)\}_{i=1}^{N_s}$, where $\lambda_i$ and $\phi_i$ are the longitude and latitude of station, $v_i$ is the observed annual mean wind speed, and $N_s$ is the total number of stations.

A regular latitude, longitude grid is generated over the study region. Let an arbitrary grid point be represented as $x_j=(\lambda_j,\phi_j)$, where $j=1,2,\ldots,M$ and $M$ is the total number of grid points. The wind speed at each grid point is estimated from the station observations using inverse distance weighting (IDW), a deterministic spatial interpolation method applied to sparse wind speed observations \cite{Ravazzani2020_IDW}. The distance between grid point $x_j$ and station $i$ is computed as
\begin{equation}
d_{ij}=\sqrt{(\lambda_j-\lambda_i)^2+(\phi_j-\phi_i)^2}.
\end{equation}

The corresponding IDW weight is defined as
\begin{equation}
w_{ij}=\frac{1}{d_{ij}^{z}+\epsilon},
\end{equation}
where $z$ is the distance decay parameter and $\epsilon$ is a small numerical constant used to avoid division by zero. The interpolated wind speed at grid point $x_j$ is then computed as
\begin{equation}
\hat{v}(x_j)=\frac{\sum_{i=1}^{N_s}w_{ij}v_i}{\sum_{i=1}^{N_s}w_{ij}}.
\end{equation}

If a grid point coincides with a station location, the observed station value is assigned directly. The resulting interpolated wind field is used as the spatial hazard layer for assigning wind exposure to geographically mapped buses and transmission corridors.

\subsection{Grid Embedding within a Geographical Area}

In this analysis, electric grid is represented as a graph $\mathcal{G}=(\mathcal{B},\mathcal{L})$, where $\mathcal{B} =\{1,2,\ldots,N_b\}$ is the set of buses and $\mathcal{L}=\{1,2,\ldots,N_l\}$ is the set of transmission branches. The IEEE 118 bus system is considered as the representative electric grid for this article \cite{ICSEG_IEEE118}. Since the standard IEEE 118 bus test system does not provide geographic coordinates, the network was first represented using cartesian coordinates form representing Bus 1 as reference origin and using relative coordinate for others buses. A control point based affine coordinate transformation was used to place the test grid (i.e., IEEE 118 bus system) within the Wyoming map following the standard approach in geometric alignment \cite{Goshtasby1986_ImageRegistration,Szeliski2006_ImageAlignment}.
Let $\mathbf{p}_i = \begin{pmatrix} x_i \\ y_i \end{pmatrix}$ denote the original Cartesian coordinate of bus $i$, and let $\mathbf{q}_i = \begin{pmatrix} \lambda_i \\ \phi_i \end{pmatrix}$ denote its desired mapped longitude and latitude in the region. Three anchor buses are used (Bus 1, Bus 59 and Bus 11) in our case and  respective transformation matrix $\mathbf{T}$ was obtained relating the cartesian coordinate to its longitude and latitude values. After $\mathbf{T}$ was determined, the transformation was applied to all IEEE 118 buses as $\mathbf{q}_i = \mathbf{T} \mathbf{p}_i$ \cite{Szeliski2006_ImageAlignment} with appropriate scaling, while embedding the system within the Wyoming geographic boundary.

\subsection{Wind Speed Modeling}

Wind speed associated with straight line wind events can be considered as a Weibull random variable which is a common practice in wind engineering and power system studies \cite{Seguro2000_Weibull,Karki2006_WindReliability}. The Weibull distribution can be used for representation of both moderate and extreme wind conditions through appropriate selection of the probability density function (PDF) parameters. The PDF associated with the straight line wind speed is expressed as \cite{celik2004statistical,lun2000study}: 
\begin{equation}
f_{V_i}(v;k_i,c_i) = \frac{k}{c}\left(\frac{v}{c}\right)^{k-1}
\exp\left[-\left(\frac{v}{c}\right)^k\right],
\end{equation}
where, $k$ is the dimensionless shape parameter, and $c$ is the scale parameter representing the characteristic wind speed.  In wind speed modeling applications, the Weibull shape parameter typically ranges from 1 to 3 i.e., $k \in (1,3]$ \cite{Safari2022_Weibull}. In this study, the shape parameter is selected as $k=2$, following previous wind power evaluation work over Wyoming \cite{Lin2003_WyomingWind}. The scale parameter $c$ is estimated for various locations in Wyoming (associated with different grid component at different location ) using shape parameter and  the annual mean wind speed obtained from the interpolated Wyoming wind exposure map. For a Weibull distributed wind speed, the mean wind speed $\mu_i$ at bus $i$ is related to $c_i$ and $k$ as \cite{Seguro2000_Weibull}:
\begin{equation}
\mu_i = c_i \Gamma\left(1+\frac{1}{k}\right)
\end{equation}
where $\Gamma(\cdot)$ is the gamma function. Thus, the bus specific scale parameter is calculated as:
\begin{equation}
c_i = \frac{\mu_i}{\Gamma\left(1+\frac{1}{k}\right)}.
\end{equation}

\subsection{Component Failure Modeling}

Power system failure modeling under extreme events are generally done at component level whereas some extends to system level approaches. Component level failure modeling focuses on representing the probabilistic failure behavior of individual power system elements based on hazard intensity and structural vulnerability, whereas system level modeling evaluates the collective response of interconnected components to such failures.

Fragility curves are widely used to represent the vulnerability and failure behavior of power system components subjected to stochastic weather conditions \cite{li2019exploiting,Panteli2016_Fragility}. Such curves can be developed using several approaches, including statistical analysis of extensive historical failure data, controlled experimental testing, evaluation of design standards, expert judgment, or combinations of these methods \cite{Panteli2016_Fragility}. In contrast, analytical methods are commonly employed when sufficient historical failure data are unavailable, formulating fragility estimation through physics based or probabilistic modeling techniques as applied in this research. The probability of transmission tower failure as a function of wind speed is modeled as \cite{baker2015efficient}: 
\begin{equation}
p_{\text{logn}}(v) = \Phi \left( \frac{\ln(v) - \ln(v_{50})}{\beta} \right)
\end{equation}
where $v_{50}$ denotes the wind speed corresponding to a failure probability of 0.5,  $\beta$ controls the dispersion (i.e., steepness) of the fragility curve,  and $\Phi(\cdot)$ represents the standard normal cumulative distribution function (CDF). 
Since wind speed is modeled as a random variable, the overall wind induced failure probability of a component is obtained by accumulating the contribution of all possible wind speed values\cite{Greco2025_WindRisk}. The component failure probability at bus $i$ is obtained by weighting the fragility function with the bus specific Weibull PDF:
\begin{equation}
P_{F,i}
=
\int_{0}^{\infty}
p_{\text{logn}}(v)
f_{V_i}(v;k_i,c_i)
\,dv,
\end{equation}
where $f_{V_i}(v;k_i,c_i)$ is the Weibull PDF and $P_{F,i}$ is the wind induced component failure probability.

\subsection{Line Failure Modeling}

After estimating the component level wind induced failure probabilities, each transmission line is treated as a corridor connecting its two terminal buses. Let $\mathcal{C}_{\ell}$ denote the set of modeled components associated with corridor $\ell$. The corridor is considered failed if at least one component in $\mathcal{C}_{\ell}$ fails. Therefore, the corridor failure probability according to \cite{Greco2025_WindRisk} is expressed as  :
\begin{equation}
P_{\text{corr},\ell}
=
1-\prod_{i\in\mathcal{C}_{\ell}}
\left(1-P_{F,i}\right),
\end{equation}
where $P_{F,i}$ is the wind induced failure probability of component $i$. If each corridor is approximated using the failure probabilities of its two terminal buses. For a line $\ell=(m,n)$, this becomes
\begin{equation}
P_{\text{corr},\ell}
=
1-\left(1-P_{F,m}\right)\left(1-P_{F,n}\right),
\end{equation}
where $P_{F,m}$ and $P_{F,n}$ are the component failure probabilities at the terminal buses of line $\ell$.

\section{Simulations and Results}

This section presents the simulation studies used to evaluate the proposed framework on the Wyoming mapped IEEE 118 bus system. The experiments quantify the observations from spatial wind exposure to component level line failure, Monte Carlo contingency generation, post-contingency power flow redistribution, cascading simulation, and finally reliability analysis. These studies are performed to identify wind vulnerable transmission lines, assess the extent to which initial wind induced outages are amplified by cascading propagation, and determine the resulting system fragmentation consequences, loss of load and reliability parameters.

\subsection{Wind Exposure and Transmission Grid Overlay}
\begin{figure}[h]
    \centering
    \includegraphics[width=\linewidth]{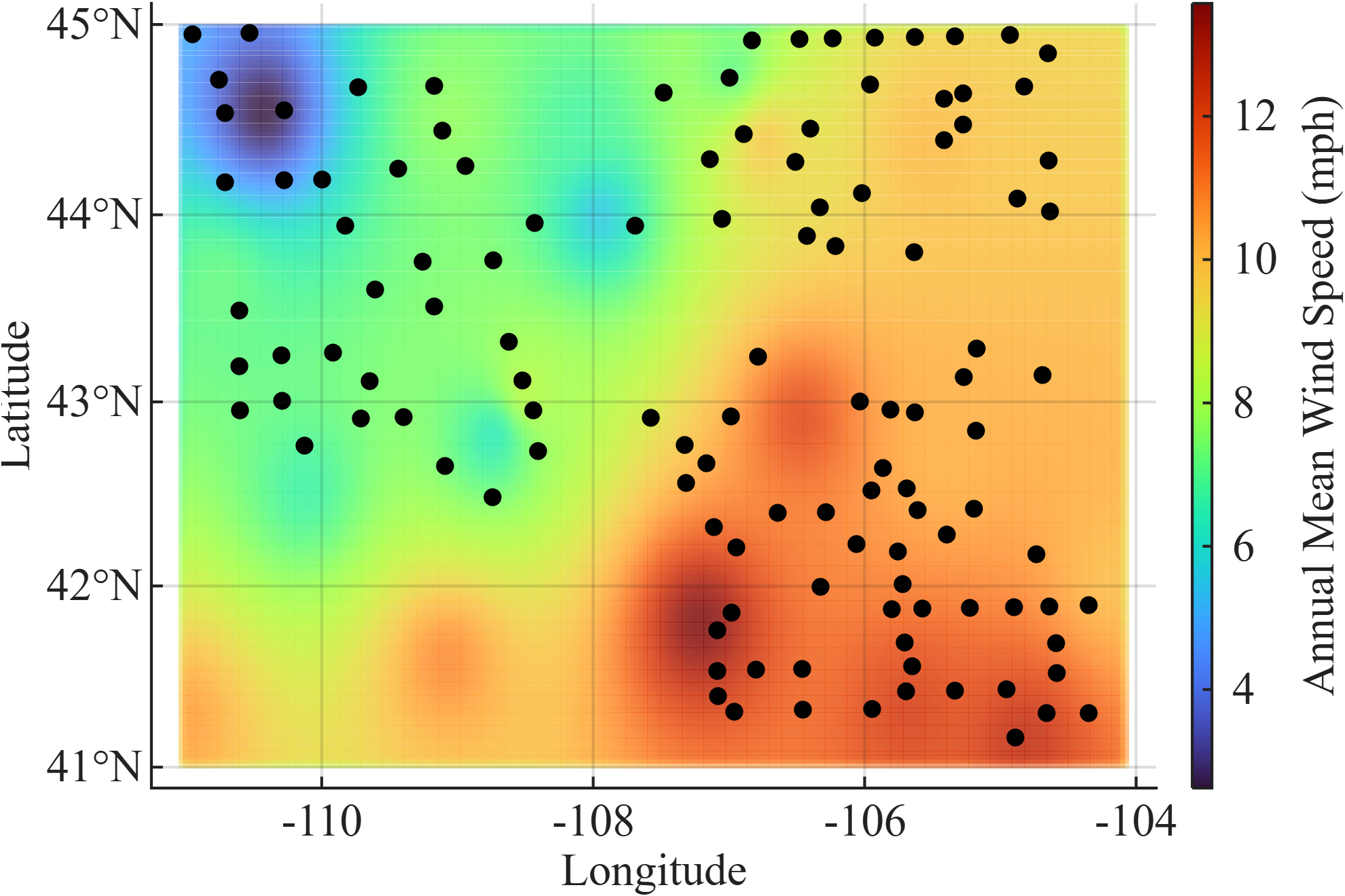}
    \caption{IEEE 118 bus network overlaid on the Wyoming wind exposure map.}
    
    \label{IEE118_WY}
\end{figure}

\begin{figure}[h]
    \centering
    \includegraphics[width=\linewidth]{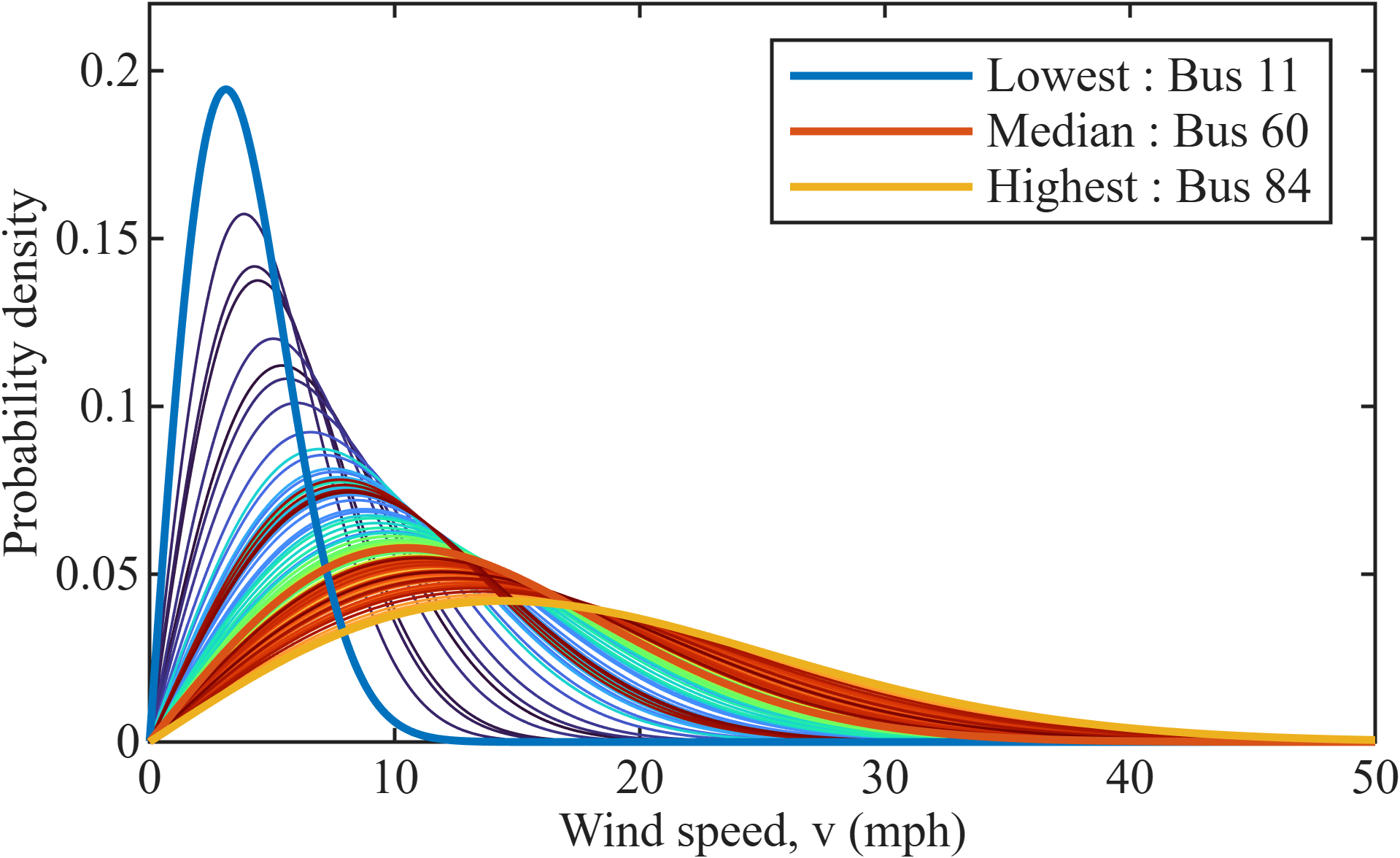}
    \caption{Wind speed probability density functions (Weibull) for buses in IEEE 118 bus system.}    
    \label{fig:weibull_pdfs}
\end{figure}

For the Wyoming case study, annual mean wind speed observations from 19 National Weather Service (NWS) stations were used to construct the spatial wind exposure field \cite{Lin2003_WyomingWind}. The IDW interpolation method described in Section II-A was applied over the Wyoming boundary using a distance decay parameter of $z=2$, giving greater influence to nearby stations than distant stations. 
This spatial hazard field was integrated with the transmission network i.e., the IEEE 118 bus system by embedding the grid within the Wyoming boundary using the affine coordinate transformation described in Section II-B. The overlaid network is shown in Fig.~\ref{IEE118_WY}.  
The annual mean wind speed obtained from the interpolated Wyoming wind exposure map was assigned to each geographically mapped bus of the IEEE 118 bus system. Using the Weibull wind speed model described in Section II-C, the shape parameter was fixed at $k=2$, while the scale parameter $c_i$ was computed separately for each bus based on its assigned annual mean wind speed. This produced bus specific wind speed probability distributions that reflect the spatial variability of wind exposure across the Wyoming mapped IEEE 118 bus network. The resulting Weibull probability density functions for all 118 buses are shown in Fig.~\ref{fig:weibull_pdfs}. These bus specific wind speed distributions are subsequently used as probabilistic hazard inputs for the component level fragility and line outage probability calculations.

\subsection{Line Failure Probabilities and Monte Carlo Contingency List}

The wind induced failure probability of each transmission tower was calculated based on the fragility curves of the towers and the corresponding wind speed distribution at the respective tower locations as described in Section II-D,E. The bus level tower failure probabilities were then used to estimate the failure probability of each transmission line corridor. 
After estimating the failure probability of each transmission line, Monte Carlo simulation was used to generate wind induced contingency scenarios. In each trial, line $\ell$ was modeled as an independent Bernoulli random variable with outage probability $P_{\text{corr},\ell}$. A uniformly distributed random number was sampled for each line and compared with its corresponding outage probability; the line was marked outaged when the sampled value was less than $P_{\text{corr},\ell}$. 

Only non-empty outage scenarios were retained in the final contingency list. A total of 20,000 contingency scenarios were generated. The resulting contingency distribution is summarized in Table~\ref{tab:mc_contingency_distribution}.
\begin{table}[H]
\centering
\caption{Distribution of Monte Carlo generated wind induced contingencies}
\label{tab:mc_contingency_distribution}
\begin{tabular}{lcc}
\hline
Contingency type & Count & Percentage (\%) \\
\hline
Single-branch failures & 14740 & 73.70 \\
Double-branch failures & 4221 & 21.11 \\
Triple-branch failures & 900 & 4.50 \\
Four or more simultaneous branch failures & 139 & 0.70 \\
\hline
Total & 20000 & 100.00 \\
\hline
\end{tabular}
\end{table}

\subsection{Wind induced Transmission Line Vulnerability}
\begin{figure}[h]
    \centering
    \includegraphics[width=\linewidth]{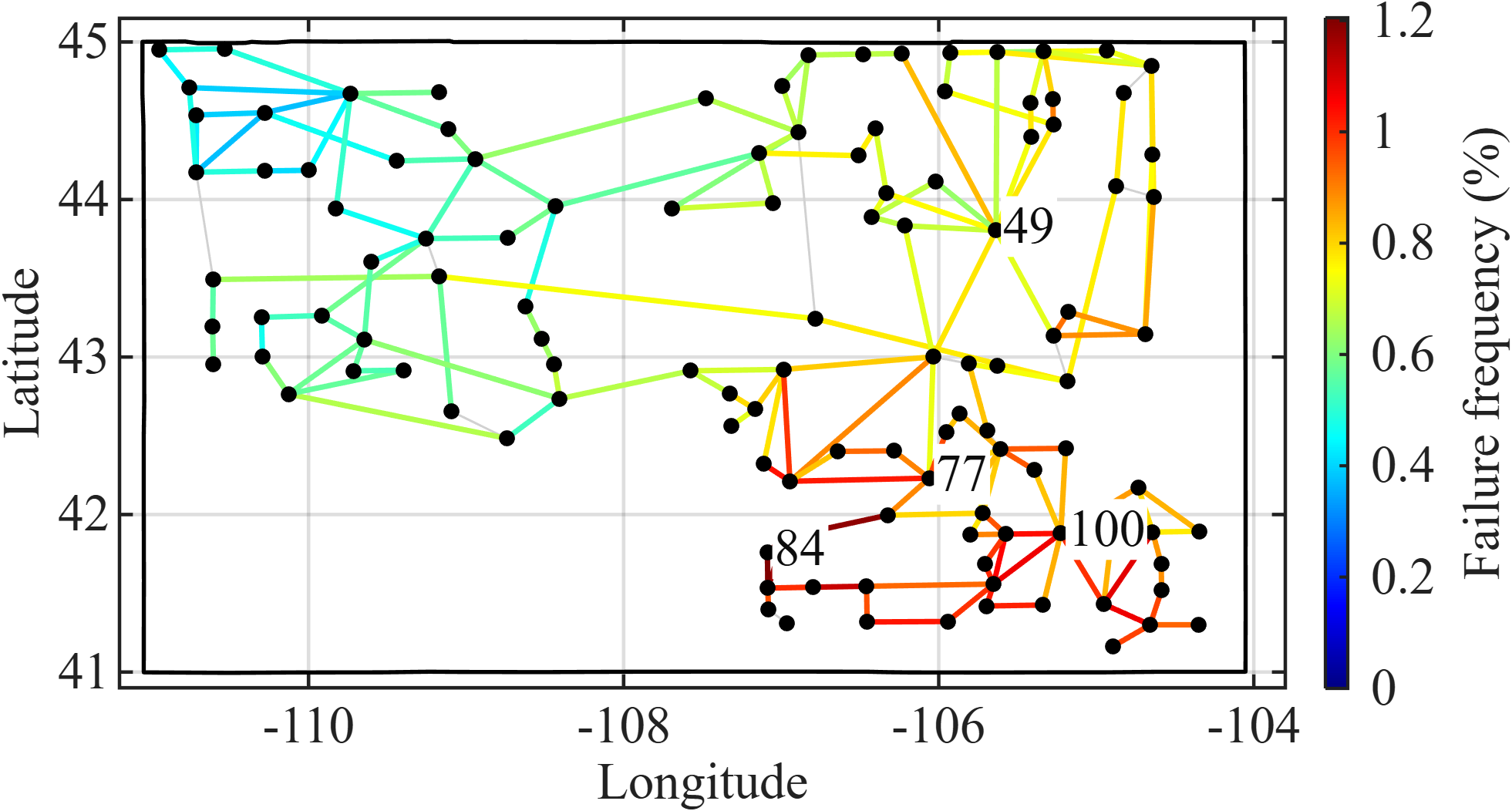}
    \caption{Transmission line vulnerability under wind exposure.}
    \label{fig6}
\end{figure}

The wind induced transmission line vulnerability map is presented in Fig.~\ref{fig6}. The figure was developed from the Monte Carlo contingency set by evaluating the occurrence frequency of each transmission line subjected to wind induced outage.
The results show a spatially nonuniform risk distribution, with higher failure probabilities concentrated in the southern portion of the mapped network, where the connected buses are exposed to stronger wind conditions. Consequently, these corridors represent more vulnerable network elements under straight line wind loading and may require prioritized consideration in grid planning, monitoring, and operational risk mitigation. Note that transformer branches and short span substation interconnections were not included in the wind induced failure frequency calculation, since these components are typically located within substation facilities, may be gas insulated, or have limited exposed conductor span \cite{kalaga2016design}. These non-modeled branches are shown in gray in Fig.~\ref{fig6} only to preserve the network topology.

\subsection{Cascading Analysis}

Cascading analysis was performed to evaluate the propagation of initial wind induced transmission line outages through the network. The standard MATPOWER IEEE 118 bus case does not provide conductor  ratings, therefore, synthetic branch flow limits were assigned based on the base case DC power flow solution \cite{schweitzer2018mathematical}. These limits serve as relative overload screening thresholds for cascading failure analysis rather than utility specified conductor ratings. For each initiating wind induced outage, the affected line was removed and a DC power flow was solved for the remaining network. The active power flow on each in-service transmission line was compared to its line rating and identified overload lines were further tripped off accordingly. Iterative power flows process continues until no additional overloads were identified.

The severity of the cascading events was evaluated using the cascading indices:  failure progression index (FPI), network fragmentation index (NFI), weighted impact metric (WIM), load shedding impact (LSI) and cascading failure impact metric (CFIM) proposed in \cite{chaudhary2025cascading}.  Table~\ref{tab:cascading_indices} summarizes the resulting cascading failure indices. On average of 20,000 contigencies, each wind induced contingency initially removed 1.3223 transmission lines. The subsequent overload driven cascading process resulted in an additional 8.5770 line outages on average. This indicates that the initial wind induced outages were amplified by the network redistribution process, producing secondary overload trips beyond the original weather driven failures for the given network.
\begin{table}[H]
\centering
\caption{Summary of cascading failure simulations.}
\label{tab:cascading_indices}
\begin{tabular}{lc}
\hline
\textbf{Metric} & \textbf{Value} \\
\hline
Lines Failed Due to Wind& 1.3223 \\
Lines Failed Due to Cascading& 8.5770 \\
Failure Progression Index (FPI) & 0.0494 \\
Network Fragmentation Index (NFI)& 0.0615 \\
Weighted Impact Metric (WIM)& 0.0332 \\
Cascading Failure Impact Metric (CFIM) & 0.0383 \\
\hline
\end{tabular}
\end{table}

The average FPI was 0.0494, indicating that approximately 4.94\% of the initially healthy transmission line set failed due to cascading propagation. The NFI was 0.0615, showing that cascading outages produced moderate topological separation of the system. The WIM was 0.0332, meaning that the failed lines represented approximately 3.32\% of the total rating weighted transmission capacity. A CFIM of 0.0383 indicates a relatively low overall cascading failure severity, with the combined effects of failure propagation, network fragmentation, weighted line impact, and load shedding corresponding to approximately 3.83\% of the maximum normalized impact. Note that equal weights were used for the four sub-indices due to the absence of utility specific weighting preferences. 

\begin{figure}[h]
    \centering
    \includegraphics[width=\linewidth]{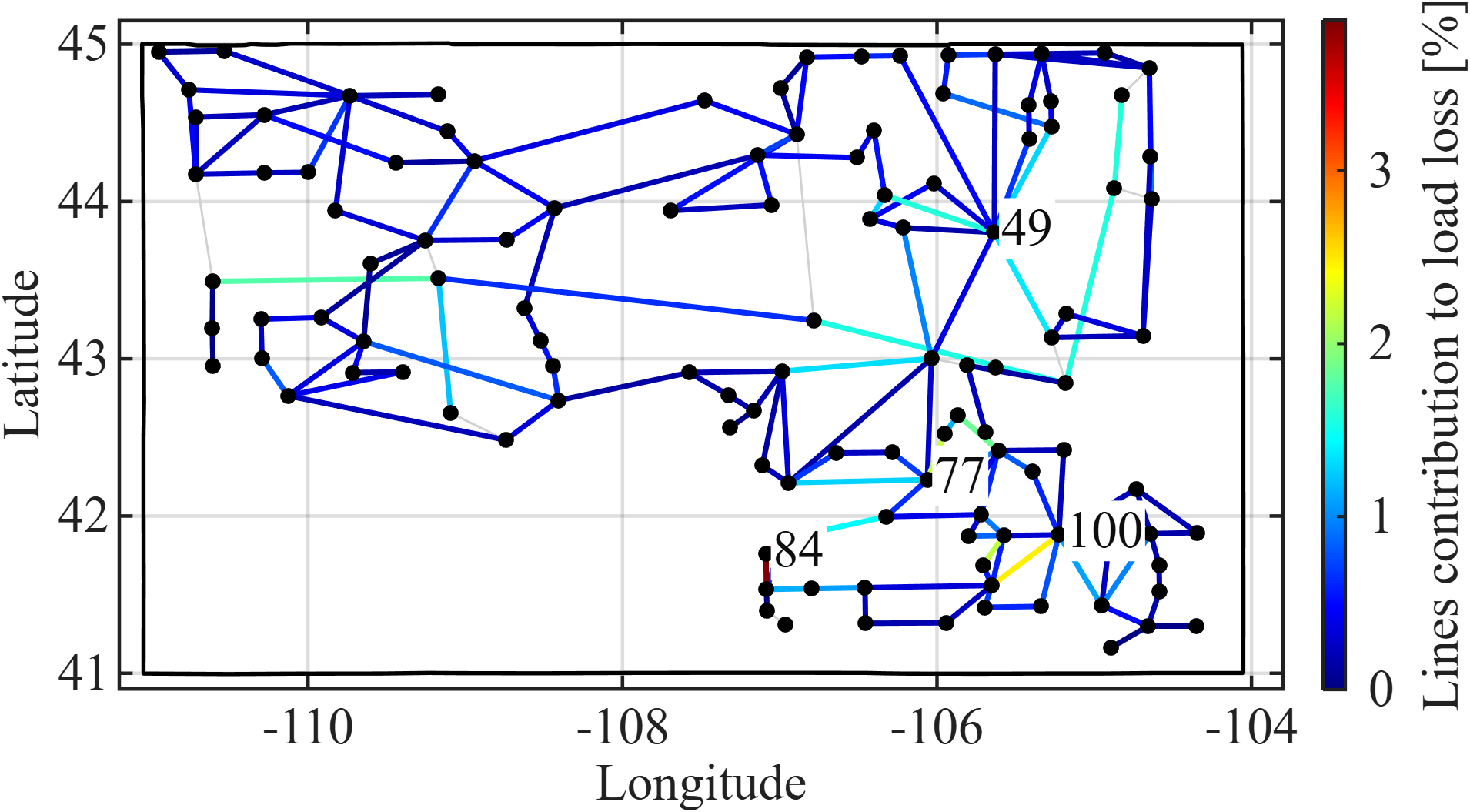}
   \caption{Critical transmission lines under wind induced cascading failures.}
    \label{critical_lines}
\end{figure}

After the cascading failure indices were evaluated, the initial wind failed lines were further ranked according to their contribution to post cascade load interruption. In this analysis, a line is considered more critical if its initial wind induced outage leads to stronger cascading propagation and larger final load not served. Therefore, the ranking does not only reflect how frequently a line is exposed to wind induced failure, but also how severely the system is affected when that line becomes an initiating outage.
The resulting critical line map is shown in Fig.~\ref{critical_lines}. Among the initial wind induced outages, line 123 connecting buses 84 and 85 was identified as the most critical corridor. When this line appeared as an initiating outage, it triggered an average of 23.42 additional cascading line trips and resulted in an average final load not served of 175.89 MW, with a maximum observed load not served of 381 MW.  The critically ranking provides a useful basis for identifying transmission corridors where hardening, monitoring, or preventive operational actions would provide the greatest reliability benefit.

\subsection{Reliability Analysis}

Following the wind induced contingency and cascading failure assessment, reliability indices were calculated from the presented scenarios in section III-B.
The reliability assessment quantifies the likelihood and severity of wind induced service interruptions of IEEE 118 bus system in Wyoming region. The reliability indices: loss of load probability (LOLP), expected demand not supplied (EDNS) and severity index presented in Table~\ref{tab:reliability_indices}  were calculated following the reliability assessment approach described in \cite{singh2018electric}.
\begin{table}[h]
\centering
\caption{Summary of Reliability indices.}
\label{tab:reliability_indices}
\begin{tabular}{lc}
\hline
\textbf{Index} & \textbf{Value} \\
\hline
Total System Load&  4242 MW \\
Maximum Load Not Served & 381 MW \\
Loss of Load Probability (LOLP) & 0.6642 \\
Expected Demand Not Supplied (EDNS) & 38.281 MW \\
Severity Index & 0.0090 p.u. \\
\hline
\end{tabular}
\end{table}

For the simulated wind induced  scenarios, the resulting LOLP was 0.6642. Thus, 66.42\% of the sampled scenarios produced nonzero load not served after cascading process. This high occurrence rate suggests that wind induced line outages frequently propagate into service interruption. The EDNS was 38.281 MW  with a total system load of 4242 MW, this corresponds to a Severity Index of 0.0090 p.u. indicating that on average unserved demand represents approximately 0.90\% of the total system demand.

\section{Conclusion and Future Works}

This paper presents a probabilistic framework for evaluating straight line wind induced failures in the Wyoming mapped IEEE 118 bus system. By evaluating component level fragility modeling to system level reliability analysis, the proposed methodology examines how these failures propagate throughout the network and how they affect grid performance. In this approach, wind speed is modeled as a random variable to estimate transmission line failure probabilities and assess the resulting cascading impacts. Moreover, the framework provides power system operators with valuable insight into how wind induced failures can compromise grid reliability. It also enables grid planners and operators to identify critical transmission lines and develop effective grid hardening strategies.\\ Further research could incorporate distributed energy resources and energy storage systems to examine their role in improving grid operations under wind induced outage conditions. The framework can also be expanded to multi-hazard scenarios, such as combined wind, wildfire, snow storms and other extreme events, to support long-term planning for weather resilient power grid operation.

\bibliographystyle{ieeetr}
\bibliography{references}

\end{document}